\documentclass[12pt,a4paper]{article}  
\usepackage{amssymb, amsmath, bbold, mathbbol}
\usepackage{hyperref}
\usepackage{color}
\usepackage{tabstackengine}

\newcommand{\be}{\begin{equation}}
\newcommand{\ee}{\end{equation}}

\def\la{\langle}
\def\ra{\rangle}
\newcommand{\blambda}{\overline{\lambda}}
\newcommand{\brho}{\overline{\rho}}
\newcommand{\Dslash}{\nabla \! \! \! \!  / \, } 
\newcommand{\bpsi}{\overline{\psi}}
\newcommand{\tr}{{\rm tr}}

\begin{document}

\begin{center}
{\LARGE \bf  Axial gravity and anomalies of fermions}
\vskip 1.2cm
Fiorenzo Bastianelli$^{\,a}$ and Matteo Broccoli$^{\,b}$ 
\vskip 1cm
$^a${\em Dipartimento di Fisica ed Astronomia, Universit{\`a} di Bologna and\\
INFN, Sezione di Bologna, via Irnerio 46, I-40126 Bologna, Italy}
\vskip .3cm
$^b${\em Max-Planck-Institut f\"ur Gravitationsphysik (Albert-Einstein-Institut)\\
 Am M\"uhlenberg 1, D-14476 Golm, Germany}
  \end{center}
\vskip .8cm

\abstract{We consider a Dirac fermion in a metric-axial-tensor (MAT) background.
By regulating it with Pauli-Villars fields we analyze and compute its full anomaly structure. 
Appropriate limits of the MAT background allows to recover the 
anomalies of Dirac and Weyl fermions in the usual curved spacetime, obtaining in 
particular the trace anomaly of a chiral fermion, which has been the object of recent analyses.  

\section{Introduction}

A  metric-axial-tensor (MAT) background for  Dirac fermions has been recently constructed 
in \cite{Bonora:2017gzz, Bonora:2018obr}, with the main purpose of addressing anomalies, 
especially in a suitable chiral limit. It  generalizes 
to curved space the approach used by Bardeen to study vector and axial couplings of  Dirac fermions 
to gauge fields and analyze their anomalies \cite{Bardeen:1969md}.
The metric-axial-tensor is defined by
\be
\hat g_{\mu\nu} = g_{\mu\nu} +\gamma^5 f_{\mu\nu}  
\ee
and induces similar axial extensions (i.e. with a $\gamma^5$ component)
to the other geometrical quantities, like the vierbein $\hat e_\mu^a$
 and the spin connection  $\hat \omega_{\mu ab}$.
A massless Dirac fermion coupled to the MAT background has a lagrangian of the form
\be
 \mathcal{L}  = -   \bpsi \gamma^a \sqrt{\hat g} \hat e_a^\mu \hat \nabla_\mu \psi
 \label{lag}
\ee
with covariant derivative
\be
\hat \nabla_\mu = \partial_\mu + \frac 14 \hat \omega_{\mu ab}\gamma^{ab} 
\ee
where $\gamma^{ab} =\frac12 [ \gamma^a,\gamma^b]$.
All quantities with a hat  contain an axial extension with $\gamma^5$ and appear
always sandwiched between the Dirac spinors $\bpsi$ and $\psi$.
For details we refer to  \cite{Bonora:2017gzz, Bonora:2018obr}, especially  appendix B of the latter. 

For our purposes it is more convenient and transparent to split the Dirac fermion $\psi$ into its two independent and Lorentz 
irreducible chiral components $\lambda$ and $\rho$ of opposite chiralities, $\psi = \lambda + \rho$. 
We use the conventions of  \cite{Bastianelli:2016nuf,Bastianelli:2018osv} for spinors and gamma matrices.  
In particular our chiral spinors satisfy 
 $ \gamma^5 \lambda =\lambda$ and $ \gamma^5 \rho =- \rho$.
Then the lagrangian takes the form
\be
 \mathcal{L} = -  \sqrt{g_+}\, \blambda \gamma^a  e^{+\mu}_a \nabla^+_\mu \lambda 
 - \sqrt{g_-}\, \brho \gamma^a e^{-\mu}_a \nabla^-_\mu \rho 
 \label{Dirac-mat}
\ee
where $g^\pm_{\mu\nu} = g_{\mu\nu} \pm f_{\mu\nu} $ are two different effective metrics, with
related compatible vierbeins, spin connections, and covariant derivatives
(which we indicate with the $\pm$ sub/superscripts).
This happens since the $\gamma^5$ matrices acting on chiral fermions  are substituted by 
the corresponding eigenvalues.
One could be more general,  allowing also for the spacetime points to have an axial extension, 
as outlined in \cite{Bonora:2018obr}, but the present formulation   is  sufficient for our purposes.

The limit $f_{\mu\nu}  = 0$  recovers the standard massless Dirac fermion in the metric $g_{\mu\nu}$.
Setting  $g_{\mu\nu}  =  f_{\mu\nu}\to \frac12  g_{\mu\nu} $  produces instead a left handed chiral
fermion $\lambda$ coupled to the final metric $g_{\mu\nu}$, while the other chirality is projected out  
(it remains coupled to the singular metric $g^-_{\mu\nu}=0$).
A less singular limit, which keeps a free propagating  right-handed fermion, is to consider 
the ``collapsing limit" \cite{Bonora:2018obr},
which consists in setting $g^+_{\mu\nu}  = g_{\mu\nu}$ and  $g^-_{\mu\nu}  = \eta_{\mu\nu}$, 
with  $\eta_{\mu\nu}$  the flat Minkowski metric.
  
The Dirac theory in the MAT background has several symmetries which may become anomalous, 
and limits on the background can be used to recover the anomalies of a chiral fermion, as in the Bardeen method.
The classical symmetries of the model are: diffeomorphisms, 
local Lorentz transformation and Weyl rescalings, 
together with their axial extensions, all of which are background symmetries since 
they act on the MAT fields as well. In addition, the model admits
global vector and axial U(1) symmetries, that rotate the spinors by a phase. 
This  global $U(1)_V\times U(1)_A$ symmetry group does 
 not act on the MAT background or any other background, 
 as we do not couple the model to the corresponding 
abelian gauge fields, though that could be done as well. 
We will review shortly these symmetries, compute systematically 
all of their anomalies, and then study the chiral limit.

To compute the anomalies we use a Pauli-Villars (PV) regularization, where the mass  term of the PV fields
is the source of the anomalies. If the mass term can be chosen to be symmetric
under a given symmetry, then there will be no anomalies in that symmetry.
Otherwise the classical breaking due to the nonsymmetric mass term sources the one-loop anomaly.
 Here we use the scheme of \cite{Diaz:1989nx,Hatsuda:1989qy},
 that casts the anomalies in the form of a regulated Fujikawa jacobian \cite{Fujikawa:1979ay, Fujikawa:1980vr},
 which allows the use of well-known heat kernel formulae for the explicit final  evaluation.
The variation of local counterterms, that parametrize the relation to different 
regularization schemes, as for example those identified by different mass terms,
can in general be employed to cancel or shift the anomalies to different sectors.
This method has already been applied successfully to several contexts in the past, 
as the case of two-dimensional $b$-$c$ systems \cite{Bastianelli:1990xn},  
which bear some analogies to the four-dimensional Weyl fermion case, 
or the more exotic model of chiral bosons  \cite{Bastianelli:1990ev}.
It is the same method used more recently 
in \cite{Bastianelli:2016nuf, Bastianelli:2018osv, Bastianelli:2019fot}
to address the trace anomalies of a Weyl fermion.
  
Before starting our systematic treatment, let us discuss the possible form of the mass term to be used 
in the PV sector. This mass term is quite arbitrary, as long as it regulates correctly the theory
by giving rise to an invertible matrix $T$ in field space, to be defined shortly.
Given this arbitrariness, one would like to choose it in the most symmetrical way as possible, 
to preserve the maximal number of symmetries at  the quantum level. 
The choice is essentially between the Dirac and Majorana masses, suitably coupled to the MAT background.
Both are legal. However choosing a Majorana mass will simplify drastically the calculations, as it allows
to maintain anomaly free the diffeomorphisms and the local Lorentz symmetry, 
together with their axial extensions. This happens as the Majorana mass keeps
a split structure for the couplings of the chiral irreducible components of the Dirac fermion
to the effective metrics $g^\pm_{\mu\nu}$, while producing 
anomalies in the Weyl and $U(1)$ symmetries and their axial extensions only.
Thus, we will choose a Majorana mass for computing the complete set of anomalies of a Dirac fermion 
in the MAT background. We will comment briefly also on the Dirac mass,
which turns out to be much less symmetric since it destroys all of the axial symmetries.
It could be employed as well, but calculations become much more cumbersome, producing more anomalies
then necessary, that eventually must be cured by adding countertems to the effective action. 
However, let us recall once more that any choice of the PV mass term is valid, as  local counterterms can
 be added to the effective action to recover the same final result, independently 
 of the regularization scheme adopted.
This arbitrariness is a general feature of the renormalization process of QFTs.

Now, let us briefly describe our method of calculation.   A lagrangian for the fields $\varphi$ 
 \be 
{\cal L} = \frac12 \varphi^T T {\cal O} \varphi  \;,
\ee
which is invariant under a linear symmetry $\delta \varphi = K\varphi $, that may act also on the background fields
contained in $T$ and $\cal O$, is regulated by  PV fields $\phi$ with  lagrangian  
\be 
{\cal L}_{\scriptscriptstyle PV} =  \frac12 \phi^T T  {\cal O} \phi +\frac12 M \phi^T T \phi 
\label{PV-l}
\ee 
where $M$ is a mass parameter which identifies the mass matrix $T$. The latter 
permits the explicit identification of the differential operator ${\cal O}$.
In fermionic theories the operator ${\cal O}^2$ appears as the regulator.
Indeed, one may verify that the non-invariance of the mass term under the symmetry of the massless action 
$\delta \phi = K \phi $ 
produces an anomalous variation of the regulated effective action $\Gamma$, that survives in the ${M \to \infty}$ limit.
In our hypercondensed  notation\footnote{The sum over (suppressed) indices includes spacetime integration,  so that
the lagrangian identifies directly the action.} it reads 
\begin{align}
i\delta \Gamma & =i \la\delta S\ra  =  \lim_{M \to \infty} \ i M \la \phi^T (TK +\frac12 \delta T) \phi \ra
= - \lim_{M \to \infty}  
{\rm Tr} \biggl [\biggl (K + \frac12 T^{-1} \delta T \biggr ) \biggl ( 
1+ \frac{\cal O}{M} \biggr )^{\!\!\! -1} \biggr ] 
\cr
&=
- \lim_{M \to \infty}  
{\rm Tr} \biggl [\biggl (K + \frac12 T^{-1} \delta T + \frac12 \frac{\delta {\cal O}}{M} \biggr ) 
\biggl ( 1- \frac{{\cal O}^2}{M^2} \biggr )^{\!\!\! -1} \biggr ] \;.
\end{align}
The function of the regulator ${\cal O}^2$ inside the trace
can now be substituted with an exponential function, that gives an equivalent 
regularization and produces the same anomaly, 
so that one finds the Fujikawa-like formula
\be
i\delta \Gamma =i \la\delta S\ra  =
- \lim_{M \to \infty}   {\rm Tr} [J e^{ i \frac{  {\cal O}^2}{M^2}}] 
\label{8}
\ee
where  $J=K + \frac12 T^{-1} \delta T + \frac12 \frac{\delta {\cal O}}{M} $ may be 
identified as the infinitesimal part of a Fujikawa jacobian.  
As described, it is entirely due to the non invariance of the PV mass term.
We use a factor of $i$ in the exponential to stress that we employ a minkowskian time in the heat kernel.
Now, heat kernel formulae may be directly applied.  In particular, we need the heat kernel coefficient 
$a_2({\cal O}^2)$, that we indicate in the notation of appendix B of \cite{Bastianelli:2018osv}.
Details on this PV scheme may be found in \cite{Diaz:1989nx,Hatsuda:1989qy}, 
and recapitulated in  \cite{Bastianelli:2016nuf, Bastianelli:2018osv} as well.

\section{Majorana mass}
In this section we regulate the Dirac theory in the MAT background with PV fields with a lagrangian of the same form as \eqref{Dirac-mat},
but augmented by a Majorana mass,  that we choose to be coupled as
  \be
  \begin{aligned} 
\Delta_{\scriptscriptstyle M}  {\cal L}
= \frac{M}{2}   \sqrt{g_+}\, ( \lambda^T C \lambda + h.c.) +  \frac{M}{2}  \sqrt{g_-}\,( \rho^T C \rho + h.c.)  
     \label{majorana-mass}
   \end{aligned}
 \ee
where $h.c.$ denotes hermitian conjugation and $C$ is the charge conjugation matrix 
that satisfies $C\gamma^a C^{-1}= - \gamma^{a T}$.
For notational simplicity we use the same symbols for the PV fields and the original variables,
since no confusion can arise in the following.
The advantage of this specific mass term is that it is invariant under diffeomorphisms,  and thus guarantees 
the absence of gravitational anomalies \cite{AlvarezGaume:1983ig} (the stress tensor remains covariantly conserved). 
In fact, inspection of the action shows that this symmetry can be extended 
to the axial diffeomorphisms as well, guaranteeing the covariant conservation of a corresponding axial stress tensor.
Let us elaborate more extensively on this point.
The usual change of coordinates $x^\mu \to x^\mu -\xi^\mu(x)$ induce the standard transformation law
on the fields as generated by the Lie derivative ${\cal L}_\xi$
\be
\begin{aligned} 
\delta e^{\pm a}_\mu &= \xi^\nu \partial_\nu e^{\pm a}_\mu + (\partial_\mu \xi^\nu)    e^{\pm a}_\nu
\equiv {\cal L}_{\xi} e^{\pm a}_\mu \\
\delta \psi &= \xi^\mu \partial_\mu \psi \equiv {\cal L}_\xi \psi\;.
\end{aligned}
\ee
However, one can define chiral
transformation rules that leave the entire massive action invariant.
 One may define left infinitesimal diffeomorphisms generated by a vector field $\xi_+^\mu(x)$
\be
\begin{aligned} 
\delta e^{+ a}_\mu &=  {\cal L}_{\xi_+} e^{+ a}_\mu \\
\delta \lambda &=  {\cal L}_{\xi_+} \lambda\\ 
\delta e^{- a}_\mu &=  0 \\
\delta \rho &=  0
\end{aligned}
\ee
and right infinitesimal diffeomorphisms generated by a vector field $\xi_-^\mu(x)$
\be
\begin{aligned} 
\delta e^{+ a}_\mu &=  0 \\
\delta \lambda &=  0 \\
\delta e^{- a}_\mu &=  {\cal L}_{\xi_-} e^{- a}_\mu \\ 
\delta \rho &=   {\cal L}_{\xi_-} \rho  \;.
\end{aligned}
\ee
It is only the sum with local parameters identified, i.e. with  $\xi^\mu =\xi_+^\mu =\xi_-^\mu$, that
plays the role of the geometrical transformation induced by the translation of the 
spacetime point $x^\mu$ described above. 
Nevertheless, they are independent symmetries of the massless and massive actions.
They acquire a clear geometrical meaning once the spacetime point $x^\mu$ is extended  to have
an axial partner \cite{Bonora:2018obr}, but we do not need to do that for the scope of the present  
investigation. These symmetries
imply that the stress tensor and its axial partner satisfy suitable covariant conservation laws.
Invariance of the mass term, and thus invariance of the full PV action,  implies that these symmetries are not 
anomalous at the quantum level.

Similarly, the action and the mass term are invariant under the local Lorentz symmetries 
that act independently on the $+$ and $-$ sector. 
On the $+$ sector (the left-handed sector) the  left-handed  local Lorentz symmetry acts 
 nontrivially by
   \be
  \begin{aligned} 
  \delta e^{+ a}_\mu &=   \omega^{+a}{}_b e^{+ b}_\mu \cr
  \delta \lambda  & =  \frac14 \omega^+_{ab} \gamma^{ab}\lambda\cr
   \delta e^{- a}_\mu &=  0 \cr
  \delta \rho  & = 0
\label{l-lL}
   \end{aligned}
 \ee
 where $\omega^+_{ab}= - \omega^+_{ba}$ are local parameters. 
  Similarly, on the right sector one has  
\be
  \begin{aligned} 
    \delta e^{+ a}_\mu &=  0 \cr
    \delta \lambda  & =  0 \cr
        \delta e^{- a}_\mu &=   \omega^{-a}{}_b e^{- b}_\mu \cr
 \delta \rho  & =  \frac14 \omega^-_{ab} \gamma^{ab}\rho \;.
\label{r-lL}
   \end{aligned}
 \ee
Evidently, these are full symmetries of the total PV lagrangian, including the mass term. 
The invariance of the regulating fields guarantees that the 
stress tensor and its axial companion remain symmetric at the quantum level.
 
 The only possible anomalies appear in the Weyl and axial Weyl symmetries, and in the vector  and axial $U(1)$ symmetries.
 It is again more convenient to consider their $\pm$ linear combinations, that act separately on the chiral sectors of the theory.
  The infinitesimal Weyl symmetries are defined by
 \be 
\begin{aligned}
\delta e_\mu^{\pm a}  & = \sigma^\pm e_\mu^{\pm a} \cr
\delta \lambda  & = -\frac32 \sigma^+ \lambda \cr
\delta \rho  & = -\frac32 \sigma^- \rho
\end{aligned}
\label{Weyl-tra-rules}
\ee
where $\sigma^\pm$ are the two independent Weyl local parameters.
The mass terms breaks them explicitly 
  \be
\delta \Delta_{\scriptscriptstyle M}  {\cal L}
= \sigma^+ \frac{M}{2}   \sqrt{g_+}\, ( \lambda^T C \lambda + h.c.) + \sigma^- \frac{M}{2}  \sqrt{g_-}\,( \rho^T C \rho + h.c.)  
 \ee
 causing anomalies to appear.
      For the global $U(1)_L\times U(1)_R$ symmetries, with independent infinitesimal parameters $\alpha^\pm$,
    we have the transformation rules
  \be 
\begin{aligned}
\delta \lambda  & = i\alpha^+ \lambda \cr
\delta \rho  & = i\alpha^- \rho \cr
\end{aligned}
\label{u1-tra-rules}
\ee    
and the PV mass term is again responsible for their breaking
 \be
\delta \Delta_{\scriptscriptstyle M} {\cal L}
= i\alpha^+ M   \sqrt{g_+}\, ( \lambda^T C \lambda - h.c.) + i \alpha^- M \sqrt{g_-}\,( \rho^T C \rho - h.c.)  \;.
 \ee

Before computing the anomalies, let us cast the lagrangian with the Dirac mass term 
using the  Dirac basis of spinors $\psi$ and $\psi_c$, so to recognize the operators in  
\eqref{PV-l} and identify our regulator ${\cal O}^2$. 
 We prefer to use $\psi_c = C^{-1} \bpsi^T$ rather than $\bpsi$, as the former has the same index structure of   
 $\psi$, and thus lives in the same spinor space.
 The massless lagrangian \eqref{lag} with the addition of the Dirac mass term
\eqref{majorana-mass}  fixes the PV lagrangian
 \be
 \mathcal{L}_{PV}   = \frac12 \psi_c^T C  \sqrt{\bar{\hat{g}}} \hat{\Dslash}  \psi
+ \frac12 \psi^T C  \sqrt{\hat{g}} \bar{\hat{\Dslash}} \psi_c
+ \frac{M}{2} ( \psi^T \sqrt{\hat g} C \psi +\psi_c^T \sqrt{\bar{\hat{g}}}C \psi_c)
\ee
 where a bar indicates a sign change in the axial extension 
(e.g. $\bar {\hat g}_{\mu\nu} = g_{\mu\nu} -\gamma^5 f_{\mu\nu}$)
and $\hat{\Dslash} = \gamma^a  \hat e_a^\mu \hat \nabla_\mu$, 
 so that on the field basis  $ \phi = \left(    \begin{array}{c} \psi\\     \psi_c    \end{array} \right) $ one finds
\begin{equation}
    T\mathcal{O} = \left(
    \begin{array}{cc}
        0 & C \sqrt{\hat{g}}  \bar{\hat{\Dslash}} \\
        C \sqrt{\bar{\hat{g}}}  {\hat{\Dslash}} & 0
    \end{array}
    \right) 
       \;, \quad
    T = \left(
    \begin{array}{cc}
\sqrt{\hat{g}} C & 0\\ 
0 &      \sqrt{\bar{\hat{g}}} C  
    \end{array}
    \right) 
    \;, \quad
    {\cal O} = \left(
    \begin{array}{cc}
    0 &  \bar{\hat{\Dslash}} \\ 
{\hat{\Dslash}}  &     0
    \end{array}
    \right) 
\end{equation}
and the regulator
\begin{equation}
    {\cal O}^2 = \left(
    \begin{array}{cc}
\bar{\hat{\Dslash}} {\hat{\Dslash}}  & 0\\ 
0 &  {\hat{\Dslash}}  \bar{\hat{\Dslash}} 
    \end{array}
    \right) \;. 
\end{equation}
Its structure is perhaps more transparent 
when the Dirac fermions are split in their  chiral parts 
\begin{equation}
    \phi = \left(
    \begin{array}{c}
    \lambda\\
    \rho \\
    \rho_c\\
    \lambda_c
    \end{array}
    \right)
\end{equation}
and one recognizes a block diagonal regulator 
\begin{equation}
    \mathcal{O}^2 
    = \left(
    \begin{array}{cccc} \mathcal{O}^2_\lambda &0  &0 &0 \\
        0& \mathcal{O}^2_\rho   &0  &0 \\
  0 & 0& \mathcal{O}^2_{\rho_c}
  &0 \\  0& 0 &0 & \mathcal{O}^2_{\lambda_c}
      \end{array}
    \right) 
\end{equation}
with entries
\be
\begin{aligned}
{\cal O}^2_\lambda &=  \Dslash^2_+ P_+ \;, \qquad  {\cal O}^2_{\lambda_c} =  \Dslash^2_+ P_- \\[2mm]
{\cal O}^2_\rho  &=  \Dslash^2_- P_- \;, \qquad  {\cal O}^2_{\rho _c} =  \Dslash^2_- P_+\;. 
\end{aligned}
\ee  
where we have used  the left/right  chiral  projectors 
$P_+=P_L = \frac{\mathbb{1}+\gamma^5}{2}$ and $ P_-= P_R = \frac{\mathbb{1}-\gamma^5}{2}$,
and denoted by $\Dslash_\pm= \gamma^a  e^{\pm\mu}_a \nabla^\pm_\mu $ 
the Dirac operators coupled to the $\pm$ effective vierbeins.

Let us now compute the anomalies. For the Weyl symmetries we get anomalies 
in the traces of the stress tensors, defined by
varying the action under the two effective vierbeins $e^\pm_{\mu a}$
\be
T_\pm^{\mu a}(x) = \frac{1}{\sqrt{g_\pm}}\frac{\delta S} {\delta e^\pm_{\mu a} (x)}\;.
\ee
In each chiral sector we use the corresponding chiral metric, and related vierbein, to perform covariant operations and take traces, and the calculation is just  a double copy of the one 
presented in \cite{Bastianelli:2016nuf}.
Recalling \eqref{8} one identifies the structure of the breaking term $J$, entirely due  
to the PV mass. Once inserted into the ``Fujikawa trace", it is computed by using 
the heat kernel coefficients $a_2({\cal O}^2)$ for the regulators ${\cal O}^2$ due to the PV fields. 
 All the steps have
been discussed in details in \cite{Bastianelli:2016nuf, Bastianelli:2018osv}, where 
in particular it was noticed that the 
term $\delta {\cal O}$ in $J$ does not contribute to the functional trace.
 In the present situation we find
for the traces of the stress tensors on the MAT background 
\be
\begin{aligned}
\la T_{+}^\mu{}_\mu \ra &= -\frac{1}{2\, (4\pi)^2} \Big [ \tr [P_+ a_{2}({\cal O}^2_\lambda)]  +  \tr[ P_- a_{2}({\cal O}^2_{\lambda_c})]\Big ]  \cr
\la T_{-}^\mu{}_\mu \ra &= -\frac{1}{2\, (4\pi)^2} \Big [ \tr [P_- a_{2}({\cal O}^2_\rho)]  +  \tr[ P_+ a_{2}({\cal O}^2_{\rho_c})]\Big ]  \cr
\end{aligned}
\label{trace-an}
\ee
where the remaining final dimensional traces are traces on the gamma matrices. 
The projectors on the regulators can be dropped, as they  get absorbed by the explicit projectors
already present in  \eqref{trace-an}.
Thus, one may use  ${\cal O}^2_\lambda = {\cal O}^2_{\lambda_c}  =\Dslash^2_+ $ and 
${\cal O}^2_\rho =  {\cal O}^2_{\rho _c} =  \Dslash^2_-$  to simplify the anomaly expressions to 
\be
\la T_{\pm}^\mu{}_\mu \ra = -\frac{1}{2\, (4\pi)^2} \tr [a_{2}(\Dslash^2_\pm)]   
\label{trace-an-2}
\ee
and one finds the following trace anomalies on the MAT background
\be
\la T_{\pm}^\mu{}_\mu \ra =
\frac{1}{720\, (4 \pi)^2} \Big ( 7 R_{\mu \nu \lambda \rho} R^{\mu \nu \lambda \rho} +8 R_{\mu \nu} R^{\mu \nu } -5 R^2 +12 \square R \Big )(g_\pm)
\label{tracepm}
\ee
where the functional dependence on  $g_\pm$ reminds that all the geometrical quantities 
and covariant operations are computed using the effective metric $g^\pm_{\mu\nu}$.
 
We now compute the $U(1)_L\times U(1)_R$ anomalies. Evidently, we are going to find again a split form.
By the Noether theorem one finds the covariantly conserved Noether currents
\be
\delta S = \int d^4 x \sqrt{g_+}\, \alpha_+ \nabla^+_\mu J_+^\mu + \int d^4 x \sqrt{g_-}\, 
\alpha_- \nabla^-_\mu J_-^\mu 
\ee
where the constants $\alpha_\pm$ in \eqref{u1-tra-rules} are extended to arbitrary functions,
with the currents taking the explicit form
$J_+^\mu= i\blambda \gamma^a  e^{+\mu}_a \lambda$ and  
$J_-^\mu= i\brho \gamma^a  e^{-\mu}_a \rho$. 
We compute their anomalies with the PV regularization and find
\be
\begin{aligned}
& \nabla^+_\mu \la J^\mu_+ \ra =  \frac{ i }{(4\pi)^2}  \Big [\tr  [P_+ a_{2}({\cal O}^2_\lambda)]  -\tr [P_- a_{2} ({\cal O}^2_{\lambda_c})]\Big ]
\cr 
& \nabla^-_\mu \la J^\mu_- \ra =  \frac{ i }{(4\pi)^2}  \Big [\tr  [P_- a_{2}({\cal O}^2_\rho)]  -\tr [P_+ a_{2} ({\cal O}^2_{\rho_c})]\Big ]
\end{aligned}
\label{chir-an}
\ee
that once more can be simplified to  
\be
 \nabla^\pm_\mu \la J^\mu_\pm \ra =  \pm \frac{ i }{(4\pi)^2}  \tr  [\gamma^5\, a_{2}(\Dslash^2_\pm)]  \;.
\ee
Their evaluation in terms of the heat kernel coefficients  produces anomalies proportional 
to the Pontryagin density of the effective metrics
\be
\nabla^\pm_\mu \la J^\mu_\pm \ra =  \mp \frac{1}{48\, (4\pi)^2} \sqrt{g_\pm}
\epsilon_{\alpha\beta\gamma\delta} R_{\mu \nu}{}^{\alpha\beta}R^{\mu \nu \gamma\delta} (g_\pm)\;.
\label{chiral-an} 
\ee

Eqs. \eqref{tracepm} and  \eqref{chiral-an} are our final results for the anomalies of a Dirac fermion on a MAT background. All other symmetries are anomaly free.

We have evaluated these anomalies using traces with chiral projectors of the heat kernel 
coefficient $a_2(\Dslash^2)$, associated to the covariant square of the Dirac operator 
in a background metric $g_{\mu\nu}$. For completeness, we list this coefficient and related traces 
\begin{align} 
a_{2}(\Dslash^2)  = &\ \frac{1}{180} (R_{\mu \nu \rho\sigma }R^{\mu \nu \rho\sigma }
-R_{\mu \nu}R^{\mu \nu})+  \frac{1}{288}R^2  -\frac{1}{120} \square  R  +\frac{1}{192} 
{\cal R}_{\mu\nu} {\cal R}^{\mu\nu}
\label{hkc}
\\
 \tr [P_\pm a_{2}(\Dslash^2)]  = 
& -\frac{1}{720} ( 7 R_{\mu \nu \rho\sigma } R^{\mu \nu\rho\sigma } +8 R_{\mu \nu} R^{\mu \nu } -5 R^2 +12 \square R) 
\cr &
\pm \frac{i}{96} \sqrt{g} \epsilon_{\alpha\beta\gamma\delta} R_{\mu \nu}{}^{\alpha\beta}R^{\mu \nu \gamma\delta} 
\end{align} 
where ${\cal R}_{\mu\nu} = R_{\mu\nu a b }\gamma^{ab}$.
One may deduce them from \cite{DeWitt:1965jb, Vassilevich:2003xt}, for example. 
They are useful in studying intermediate results leading to the evaluation 
of  \eqref{trace-an} and \eqref{chir-an},
and  have appeared in the anomaly context already in \cite{Christensen:1978gi,Christensen:1978md}. 

\section{Limits of the MAT background}

We now discuss the limits on the MAT background to recover the usual theories of Dirac and Weyl 
fermions in a curved spacetime and their anomalies.

Setting $f_{\mu\nu}=0$ reproduces the standard coupling of a massless Dirac fermion to a curved 
background and
corresponds to identify the two effective metrics $g^+_{\mu\nu}= g^-_{\mu\nu}$. 
The final stress tensor becomes the sum of the two chiral stress tensors, and acquires 
the sum of the two trace anomalies in \eqref{tracepm}.
Thus, one recovers the usual trace anomaly of a Dirac field
\be
\la T^\mu{}_\mu \ra =
\frac{1}{360\, (4 \pi)^2} \Big ( 7 R_{\mu \nu \rho\sigma } R^{\mu \nu \rho\sigma } +8 R_{\mu \nu} R^{\mu \nu } -5 R^2 +12 \square R\Big ) \;.
\label{trace-dirac}
\ee
Similarly, for the two U(1) symmetries, one obtains
 \be
\nabla_\mu \la J^\mu_\pm \ra =  \mp \frac{1}{48\, (4\pi)^2} \sqrt{g}
\epsilon_{\alpha\beta\gamma\delta} R_{\mu \nu}{}^{\alpha\beta}R^{\mu \nu \gamma\delta} 
\ee
 which gets translated into the  covariant conservation of the vector current 
 $J^\mu_{_V} = J^\mu_+ +J^\mu_-$,  together with
the anomalous conservation of the axial current $J^\mu_{_A} = J^\mu_+ - J^\mu_-$,
 with the well-known Pontryagin contribution \cite{Kimura:1969wj, Delbourgo:1972xb}
 \be
 \nabla_\mu \la J^\mu_{_V} \ra =  0 \;, \qquad 
 \nabla_\mu \la J^\mu_{_A} \ra =  - \frac{1}{24\, (4\pi)^2} \sqrt{g}
\epsilon_{\alpha\beta\gamma\delta} R_{\mu \nu}{}^{\alpha\beta}R^{\mu \nu \gamma\delta}  \;.
\label{chiral-an-dirac}
\ee

Let us now study the case of the Weyl fermion $\lambda$.
This is obtained by taking the collapsing limit in which the effective metric  $g^-_{\mu\nu}$ becomes flat 
($g^-_{\mu\nu}=\eta_{\mu\nu}$ and $g^+_{\mu\nu}=g_{\mu\nu}$),  so that 
the independent right-handed fermion $\rho$ decouples  completely from the background.
Therefore, only the chiral left-handed part part contributes to the stress tensor, producing
for the trace anomaly half of the result above. 
 Similarly, one finds the anomalous conservation of the $U(1)$ current $J^\mu_+$, 
the only one that remains coupled to the curved background, 
 with the expected Pontryagin contribution.
 To summarize, we find for a left-handed Weyl fermion the following anomalies
   \be
  \begin{aligned}
  \la T^\mu{}_\mu \ra &=
\frac{1}{720\, (4 \pi)^2} \Big ( 7 R_{\mu \nu \rho\sigma } R^{\mu \nu \rho\sigma } +8 R_{\mu \nu} R^{\mu \nu } -5 R^2 +12 \square R \Big ) 
\cr
\nabla_\mu \la J^\mu_+ \ra &=  - \frac{1}{48\, (4\pi)^2} \sqrt{g}
\epsilon_{\alpha\beta\gamma\delta} R_{\mu \nu}{}^{\alpha\beta}R^{\mu \nu \gamma\delta} \;.
\end{aligned}
\ee

These results confirm the absence of a Pontryagin term in the trace anomaly of a Weyl fermion, 
as calculated in \cite{Bastianelli:2016nuf} and confirmed in \cite{Frob:2019dgf}.
The Pontryagin term sits only in the chiral anomaly.

\section{Dirac mass}

In this section we wish to give a  brief description of a different regularization, 
namely the one obtained by using a Dirac mass for the PV fields. 
In a flat background a Dirac mass is given by
\begin{equation}
 - M\bar{\psi}\psi =  \frac12 M (\psi_c^T C \psi + \psi^T C \psi_c)\;.
\end{equation}
As well-known this term
breaks the $U(1)_A$ axial symmetry while maintaining the $U(1)_V$ vector symmetry.
This continues to be the case also when one tries to MAT-covariantize it. 
There are various options to couple the Dirac mass to the MAT geometry.
One may choose to use in the mass term only  the metric $g_{\mu\nu}$, without any axial extension,
so that 
\begin{equation}
  \Delta_{\scriptscriptstyle D} \mathcal{L} = - \sqrt{g} M\bar{\psi}\psi 
  = \frac{\sqrt{g}}{2} M (\psi^T_c C \psi + \psi^T C \psi_c )
  \label{dir-mass}
\end{equation}
has the virtue of preserving the vector-like diffeomorphisms and  vector-like local Lorentz transformations
on top of the $U(1)_V$ symmetry, while breaking all of their axial extensions.
It also breaks both vector and axial Weyl symmetries, 
which are then expected to be anomalous as well. 
Counterterms should eventually be introduced to achieve the equivalence 
with our previous results.
Other choices are also possible, as for example $\sqrt{g} \to \tfrac12( \sqrt{g_+} + \sqrt{g_-})$,
which shares the same property of 
 preserving the vector-like diffeomorphisms  and local Lorentz transformations.

In the following we just  wish to derive the regulators to be used  for computing 
the anomalies in this new scheme.
  We add to the lagrangian \eqref{lag} written in a symmetric form 
\be
 \mathcal{L}  =  
  \frac12 \psi_c^T C  \sqrt{\bar{\hat{g}}} \hat{\Dslash}  \psi
+ \frac12 \psi^T C  \sqrt{\hat{g}} \bar{\hat{\Dslash}} \psi_c
 \label{lag-2}
\ee
the mass term \eqref{dir-mass}, and comparing it with \eqref{PV-l}  we find
on the field basis  
$
    \phi = \left(
    \begin{array}{c}
    \psi\\
    \psi_c
    \end{array}
    \right) 
    $
\begin{equation}
    T\mathcal{O} = \left(
    \begin{array}{cc}
        0 & C \sqrt{\hat{g}}  \bar{\hat{\Dslash}} \\
        C \sqrt{\bar{\hat{g}}}  {\hat{\Dslash}} & 0
    \end{array}
    \right) 
 \, , \quad
    T = \left(
    \begin{array}{cc}
        0 & \sqrt{g}  C \\
   \sqrt{g}         C & 0
    \end{array}
    \right) \, , \quad
    \mathcal{O} =\left(
    \begin{array}{cc}
    \sqrt{\frac{\bar{\hat{g}}}{g}} \, {\hat{\Dslash}}
         & 0 \\
        0 & 
        \sqrt{\frac{\hat{g}}{g}} \, \bar{\hat{\Dslash}}
    \end{array}
    \right) \;.
\end{equation}
Thus, we get the regulator
\begin{equation}
    \mathcal{O}^2 = \left(
    \begin{array}{cc}
        \sqrt{\frac{\bar{\hat{g}}}{g}} \, {\hat{\Dslash}} \sqrt{\frac{\bar{\hat{g}}}{g}} \, {\hat{\Dslash}}
        & 0 \\
        0 &   \sqrt{\frac{\hat{g}}{g}} \, \bar{\hat{\Dslash}}  \sqrt{\frac{\hat{g}}{g}} \, \bar{\hat{\Dslash}}
    \end{array}
    \right) 
\end{equation}
with the differential operators acting on everything placed on their right hand side. 
This regulator  $\mathcal{O}^2$  is difficult to work with, but it has the virtue of being
covariant under vector diffeomorphisms, making it somewhat manageable after all. 
Its structure is again more transparent when splitting the Dirac fermion into its  chiral parts, so that on the basis 
\begin{equation}
    \phi = \left(
    \begin{array}{c}
    \lambda\\
    \rho \\
    \rho_c\\
    \lambda_c
    \end{array}
    \right)
\end{equation}
one finds a block diagonal regulator 
\begin{equation}
    \mathcal{O}^2 
    = \left(
    \begin{array}{cccc} \mathcal{O}^2_\lambda &0  &0 &0 \\
        0& \mathcal{O}^2_\rho   &0  &0 \\
  0 & 0& \mathcal{O}^2_{\rho_c}
  &0 \\  0& 0 &0 & \mathcal{O}^2_{\lambda_c}
      \end{array}
    \right) 
\end{equation}
with entries

\be
\begin{aligned}
\mathcal{O}^2_\lambda &=  \sqrt{\frac{g_-}{g}}\Dslash_- \sqrt{\frac{g_+}{g}}\Dslash_+ P_+ 
\;, \qquad
\mathcal{O}^2_\rho = \sqrt{\frac{g_+}{g}}\Dslash_+  \sqrt{\frac{g_-}{g}}\Dslash_- P_-  \\
\mathcal{O}^2_{\lambda_c}  &=  \sqrt{\frac{g_-}{g}}\Dslash_- \sqrt{\frac{g_+}{g}}\Dslash_+ P_- 
\;, \qquad
\mathcal{O}^2_{\rho_c}  =  \sqrt{\frac{g_+}{g}}\Dslash_+  \sqrt{\frac{g_-}{g}}\Dslash_- P_+  \;.
  \end{aligned}
\end{equation}
The functions $\sqrt{\frac{g_\mp}{g}}$ are scalar functions under 
the vector-like diffeomorphism, so that these regulators are covariant under that symmetry.
The projectors $P_\pm$ take just the unit value on the  corresponding chiral spinor space,
but we have kept them to remember on which space the different regulators act.
A systematic analysis of all the anomalies, including the axial gravitational anomaly, 
may be feasible in this scheme, at least  when treating
$f_{\mu\nu}$ as a perturbation.

\section{Conclusions}
We have studied the full set of anomalies of a Dirac fermion coupled to 
the MAT background formulated recently in \cite{Bonora:2017gzz,Bonora:2018obr}.
This result has allowed a rederivation of the anomalies of a Weyl fermion coupled 
to a curved spacetime, including  the trace anomaly. 
Our result for the trace anomaly agrees with the one 
 calculated in \cite{Bastianelli:2016nuf} and reproduced 
with different methods  in \cite{Frob:2019dgf}. 

These findings however are at odds with the original claim of ref. \cite{Bonora:2014qla}, 
 reconfirmed also in \cite{Bonora:2017gzz,Bonora:2018obr} 
where the notion of the MAT background was developed
precisely for the purpose of studying the anomalies.
Let us comment a bit more on this point. 
The presence of a Pontryagin term, which satisfies the consistency conditions for 
trace anomalies \cite{Bonora:1985cq} and would constitute a type-B anomaly in the classification of \cite{Deser:1993yx},
was conjectured to be a realistic possibility in \cite{Nakayama:2012gu}, see also 
\cite{Bonora:2015nqa,Nakayama:2018dig, Nakayama:2019mpz}. 
On the other hand, it is known that CFTs do not support nonlocal parity-odd terms in the correlation function 
of three stress tensors
\cite{Stanev:2012nq,Zhiboedov:2012bm}, thus hinting at the absence of such a contribution.
Our explicit calculation within the MAT background shows indeed that such terms are absent,
thereby confirming the findings of refs. \cite{Bastianelli:2016nuf} and \cite{Frob:2019dgf}.
The  analogous case of a  Weyl fermion in a gauge background has also been studied more recently
in \cite{Bastianelli:2018osv, Bastianelli:2019fot}, where it was found 
that parity-odd terms do not contribute to the trace anomaly in that context as well.

Retracing our calculation of the trace anomaly, and observing the formulae in \eqref{trace-an},  
one may notice that an imaginary  term proportional to the Pontryagin density would 
indeed arise in the trace anomalies if the contribution from the regulators of the charge conjugated 
fields were neglected. However, there is no justification for dropping those terms. 
A close analogy is given by the two-dimensional $b$-$c$ system, whose 
gravitational and trace anomalies have been computed in \cite{Bastianelli:1990xn}
with the same methods employed here. In that paper, the contributions from the regulator 
of the $c$ field and that of the  $b$ field must be added together, and they sum up to produce  
the correct final anomaly. It would not be correct to drop, say  the $b$ contribution 
to find the anomaly of the $c$ field. Said differently, the propagator of the  $b$-$c$ system 
contains information on both fields, and they cannot be split artificially.
Similarly, in the Weyl fermion case both helicities $h=\pm \frac12$ (described by the Weyl fermion and its
hermitian conjugate) circulate in the loop that produces the anomalies, 
and their contributions cannot be split in any legal way.
This is consistent with four dimensional CPT,  that requires both helicities to be present in a massless 
relativistic QFT.

A preprint has recently appeared \cite{Bonora:2019dyv},
suggesting that the methods we use for the anomaly calculations are not fit
to detect parity-odd terms in the trace anomaly.
We reject those criticisms, which we find unfounded. We find it important 
to reiterate that in principle any regularization scheme can 
be used to define a QFT, with different schemes  producing perhaps different results for the anomalies, 
that however must be related by adding local counterterms to the effective action. 
The use of massive PV fields with a Majorana mass is certainly legal for regulating Weyl fields.
 We have used it here to compute systematically
all the anomalies within the same regularization scheme, finding in particular  the correct 
and well-known consistent $U(1)$ chiral anomaly.
The Majorana mass couples the two helicities of a Weyl fermion, 
thus breaking its U(1) symmetry that indeed becomes anomalous. 
It also breaks the Weyl local scaling symmetry, thus causing a trace anomaly to appear as well.
Other regularizations may of course be used, though some of them 
might be too cumbersome to carry out effectively the calculations 
(we have illustrated briefly the case of PV fields with Dirac mass in the last section).
As a final comment on the views expressed in \cite{Bonora:2019dyv}, 
we wish to stress that the theory of a chiral fermion in 4 dimensions
may be described equivalently in two ways. One description makes use of a 
Weyl spinor and its hermitian conjugate (this is how we have proceeded in the present paper).
Alternatively, one may use a Majorana spinor. The latter has the same field content of the former, 
as it casts together the two irreps   of the Lorentz group ((1/2, 0) 
for the Weyl spinor plus its complex conjugate (0, 1/2) 
for the hermitian conjugate Weyl spinor) 
into a single spinor, making the resulting Majorana spinor reducible under the Lorentz group. 
 Lorentz invariance fixes uniquely their actions, which are totally equivalent.
  A mass term can be added in both schemes, it is the so-called Majorana mass, which has the property of  
  breaking the chiral $U(1)$ symmetry associated to the conservation of a fermion number 
 (correlated to the helicity and identified with the lepton number in neutrino applications).
  References where these matters are clearly explained are the textbooks \cite{Srednicki:2007qs, Banks:2008tpa, AlvarezGaume:2012zz}, for example.
Thus, we find unfounded the suggestion that our methods are suitable for Majorana fermions but 
  not for Weyl fermions.

\section*{Acknowledgments}
We acknowledge Loriano Bonora for useful discussions and email exchanges. MB thanks Lorenzo Casarin and Stefan Theisen for discussions. MB is supported by the International Max Planck Research School for Mathematical and Physical Aspects of Gravitation, Cosmology and Quantum Field Theory. 


\end{document}